\documentclass[11pt]{article}
\usepackage{graphicx}
\usepackage{amsmath}
\usepackage{amssymb}

\begin{document}

\title{de Sitter geodesics}

\author{{Ion I. Cot\u aescu\footnote{E-mail: i.cotaescu@e-uvt.ro}}\\
{\small \it West University of Timi\c{s}oara,}\\
{\small \it V.  P\^{a}rvan Ave.  4, RO-300223 Timi\c{s}oara, Romania}}

\maketitle

\begin{abstract}
The geodesics on the $(1+3)$-dimensional de Sitter spacetime are considered studying how their parameters are determined by the conserved quantities in the conformal Euclidean, Friedmann-Lema\^itre-Robertson-Walker, de Sitter-Painlev\'e and static local charts with Cartesian space coordinates.  Moreover, it is shown that there exist a special static  chart in which the geodesics are genuine  hyperbolas whose asymptotes are given by the conserved momentum and the associated dual momentum.
\end{abstract}


\newpage

\section{Introduction}

The free geodesic motion on the  $(1+3)$-dimensional de Sitter (dS) spacetime was considered by may authors which studied the role of the conserved quantities  \cite{cac1,cac2} or the so called dS relativity in a dS local chart where the Lorentz transformations of the $SO(1,4)$  isomety group have the same form as in special relativity \cite{dSrel}.  

Recently we proposed our own version of  relativity on anti-de Sitter \cite{CAdS1,CAdS2} and dS  \cite{CdS1,CdS2} backgrounds which solves completely the problem of the relative geodesic motion thanks to the Lorentzian isometries that can transform among themselves any local charts in which we study geodesic trajectories \cite{CAdS2,CdS1}.
Our approach is based on the classical conserved quantities given by the Killing vectors associated to isometries. With their help we marked the different local charts as fixed or mobile frames finding the parametrization of the Lorenzian isometries relating them.  

In the case of the dS spacetimes we studied the conserved quantities associated to the $SO(1,4)$ isometries  \cite{ES,CDir,CGRG} showing that  apart from the energy, momentum and angular momentum, there exists a dual momentum which plays an important role in understanding the significance of the principal classical invariant that in the flat limit gives the mass condition of special relativity \cite{CGRG,CdS1}. These conserved quantities which helped us to build the dS relativity play an important role in determining the geodesic trajectories.  Here we would like to concentrate on this problem completing thus our previous investigations with a systematic study of  the dS geodesics. 

In general, on the dS spacetime the geodesics depend on the initial condition and the conserved momentum which determine the trajectory parameters and, implicitly, the conserved quantities along geodesics.  The inverse problem we intend to study here is how the initial conditions depend on the conserved quantities when these are given. In other words, we would like to investigate how the form of the geodesic trajectories depend on these conserved quantities,  eliminating the undetermined arbitrary initial conditions.   

This goal restricts the investigation area of the present paper which remains thus a technical review rather then a major original contribution. Nevertheless, here we obtain new results concerning the properties of the above mentioned conserved quantities and their role  in determining geodesic parameters.  Moreover, we introduce a new conserved vector which offers one some technical advantages when we use exclusively Cartesian space coordinates as we proceed  in this paper.  

We start in the second section presenting the physical meaning of the classical conserved quantities related to the dS isometries and the local charts we use. In the next section we discuss the properties of these quantities in the conformal Euclidean chart and we introduce the new conserved vector that helps us to separate the contribution of the initial conditions to the geodesic equations  of this chart. The section 3 is devoted to the geodesics in comovong charts with Cartesian space coordinates, i. e. conformal Euclidean, Friedmann-Lema\^itre-Robertson-Walker (FLRW) and de Sitter-Painlev\'e (dSP) ones. The goodesics in the static and special static charts with similar space coordinates are studied in the next section. Finally, in section 6, a special attention is paid to the form of the null cones of all the local charts we consider here.  The presence of the event horizons \cite{BD} in the dSP and static charts is pointed out and briefly commented. Few concluding remarks are presented in the last section.

\section{Preliminaries}

The de Sitter spacetime $(M,g)$ is defined as a hyperboloid of radius $1/\omega$  in the five-dimensional flat spacetime $(M^5,\eta^5)$ of coordinates $z^A$  (labeled by the indices $A,\,B,...= 0,1,2,3,4$) having the pseudo-Euclidean metric $\eta^5={\rm diag}(1,-1,-1,-1,-1)$. The local charts $\{x\}$  of coordinates $x^{\mu}$ ($\alpha,\mu,\nu,...=0,1,2,3$) can be introduced on $(M,g)$ giving the set of functions $z^A(x)$ which solve the hyperboloid equation,
\begin{equation}\label{hip}
\eta^5_{AB}z^A(x) z^B(x)=-\frac{1}{\omega^2}\,,
\end{equation}
where  $\omega$ denotes the Hubble de Sitter constant since in our notations  $H$ is reserved for the energy (or Hamiltonian) operator \cite{CGRG}. 

The de Sitter isometry group is just the gauge group $G(\eta^5)=SO(1,4)$ of the embedding manifold $(M^5,\eta^5)$  that leave  invariant the metric $\eta^5$ and implicitly Eq. (\ref{hip}). Therefore, given a system of coordinates, defined by the functions $z=z(x)$, each transformation ${\mathfrak g}\in SO(1,4)$ defines the isometry $x\to x'=\phi_{\mathfrak g}(x)$ derived from the system of equations $z[\phi_{\mathfrak g}(x)]={\mathfrak g}z(x)$. For these isometries we use the canonical parametrization
\begin{equation}
{\mathfrak g}(\xi)=\exp\left(-\frac{i}{2}\,\xi^{AB}{\mathfrak S}_{AB}\right)\in SO(1,4) 
\end{equation}
with skew-symmetric parameters, $\xi^{AB}=-\xi^{BA}$,  and the covariant generators ${\mathfrak S}_{AB}$ of the fundamental representation of the $so(1,4)$ algebra carried by $M^5$. These generators have the matrix elements, 
\begin{equation}
({\mathfrak S}_{AB})^{C\,\cdot}_{\cdot\,D}=i\left(\delta^C_A\, \eta_{BD}^5
-\delta^C_B\, \eta_{AD}^5\right)\,.
\end{equation}
The principal $so(1,4)$ basis-generators with an obvious physical meaning \cite{CGRG} are the energy ${\mathfrak H}=\omega{\mathfrak S}_{04}$, angular momentum  ${\mathfrak J}_k=\frac{1}{2}\varepsilon_{kij}{\mathfrak S}_{ij}$, Lorentz boosts ${\mathfrak K}_i={\mathfrak S}_{0i}$, and a Runge-Lenz-type vector, ${\mathfrak R}_i={\mathfrak S}_{i4}$, generating rotation involving the $z^4$ axis. The effect of the $SO(1,4)$ isometries depends on the concrete coordonates we use  as we showed in Refs. \cite{CdS1}. 

The corresponding classical conserved quantities  can be derived with the help of  the  Killing vectors  $k_{(AB)}$ whose  covariant components in an arbitrary chart $\{x\}$ of  $(M,g)$ are defined as
\begin{equation}\label{KIL}
k_{(AB)\,\mu}= z_A\partial_{\mu}z_B-z_B\partial_{\mu}z_A\,, 
\end{equation}
where $z_A=\eta_{AB}z^B$. The principal conserved quantities along a timelike geodesic of a point-like particle of mass $m$ and momentum $\vec{P}$ \cite{CGRG} have the general form  ${\cal K}_{(AB)}=\omega k_{(AB)\,\mu}m u^{\mu}$ where $u^{\mu}=\frac{dx^{\mu}(s)}{ds}$ are the components of the covariant four-velocity that satisfy  $u^2=g_{\mu\nu}u^{\mu}u^{\nu}=1$. The conserved quantities with physical meaning \cite{CGRG} are, 
\begin{eqnarray}
{\mathfrak H} &\to& E=\omega  k_{(04)\,\mu}m u^{\mu}\label{consE}\\
{\mathfrak J}_i&\to& L_i= \frac{1}{2}\varepsilon_{ijk}k_{(jk)\,\mu}m u^{\mu}\\
{\mathfrak K}_i&\to& K_i=  k_{(0i)\,\mu} m u^{\mu}\\
{\mathfrak R}_i&\to& R_i=  k_{(i4)\,\mu} m u^{\mu}\,,\label{consR}
\end{eqnarray}
where $E$ is the conserved energy,  $L_i$ are the usual components of angular momentum while  $K_i$ and $R_i$  are related to  the conserved momentum, $\vec{P}$, and its associated dual momentum, $\vec{Q}$, whose components read \cite{CGRG}
\begin{equation}\label{PQKR}
P^i=-\omega(R_i+K_i)\,, \quad Q^i=\omega(K_i-R_i)\,. 
\end{equation}

In what follows we use only Cartesian space coordinates which satisfy  $z^i\propto x^i$ such that the $SO(3)$ symmetry becomes global, any quantity bearing space indices transforming under rotations as $SO(3)$ vectors and tensors. Under such circumstances, we may use the vector notation for the $SO(3)$ vectors, including the position one, $\vec{x}=(x^1,x^2,x^3)\in {\mathbb R}^3$. The norms of the conserved vectors will be denoted simply as $V=|\vec{V}|$.  

Three sets of coordinates are under consideration here: those of the conformal Euclidean chart (called here simply conformal chart) denoted by $(t_c,\vec{x}_c)$, the dSP coordinates $(t,\vec{x})$ and the static coordinates $(t_s, \vec{x}_s)$ where $t_s$ is the static time of the usual static chart $\{t_s,\vec{x}\}$ while $\vec{x}_s$ are the special static space coordinates defined in Refs. \cite{P1,P2}. All these coordinates are related as  
 \begin{eqnarray}\label{trcor}
t_c&=&-\frac{1}{\omega}\,e^{-\omega t}=-\frac{1}{\omega}\sqrt{1+\omega^2\vec{x}_s^2}\,e^{-\omega t_s}\,,\\
\vec{x}_c&=&\vec{x}e^{-\omega t}=\vec{x}_s e^{-\omega t_s}\,,
\end{eqnarray}
and can be combined for defining various local charts. In the following table
\begin{center}
\begin{tabular}{lccc}
&~~$\vec{x}_c$~~~&~~~$\vec{x}$~~~&~~~$\vec{x}_s$~~~\\
$t_c$~~&~~conformal ~~& & \\
$t$&FLRW&dSP&\\
$t_s$&&static&static special
\end{tabular}
\end{center}
we list the charts in which we have to study the geodesic equations either in covariant parametric form, $x=x(\lambda)$, or in the closed form $\vec{x}=\vec{x}(t)$.  

\section{Conserved quantities in conformal charts}

Let us start  with the {conformal} charts, $\{t_c,\vec{x}_c\}$, with the conformal time $t_c$ and Cartesian spaces coordinates $x^i_c$ ($i,j,k,...=1,2,3$), defined by the functions 
\begin{eqnarray}
z^0(x_c)&=&-\frac{1}{2\omega^2 t}\left[1-\omega^2({t}_c^2 - \vec{x}_c^2)\right]\,,
\nonumber\\
z^i(x_c)&=&-\frac{1}{\omega t}x^i_c \,, \label{Zx}\\
z^4(x_c)&=&-\frac{1}{2\omega^2 t}\left[1+\omega^2({t}_c^2 - \vec{x}_c^2)\right]\,.
\nonumber
\end{eqnarray}
These charts  cover the expanding part of $M$ for $t_c \in (-\infty,0)$
and $\vec{x}_c\in {\mathbb R}^3$ while the collapsing part is covered by
similar charts with $t_c >0$. In both these cases we have the same conformal flat line element,
\begin{equation}\label{mconf}
ds^{2}=\eta^5_{AB}dz^A(x_c)dz^B(x_c)=\frac{1}{\omega^2 {t}_c^2}\left({dt}_c^{2}-d\vec{x}_c^2\right)\,,
\end{equation}
We stress that here we restrict ourselves to consider only the expanding portion (with $t_c<0$) which is a possible  model of our expanding universe. 

In this chart  the contravariant components of the Killing vectors can be calculated according to Eq. (\ref{KIL}) as 
\begin{eqnarray}
&&k^0_{(0i)}=k^0_{(4i)}=-\omega t_c x_c^i\,,\nonumber\\
&&k^j_{(0i)}=k^j_{(4i)}+\frac{1}{\omega}\,\delta^j_i=- \omega x_c^i x_c^j+\delta^j_i\chi\,,\nonumber\\
&&k^0_{(ij)}=0\,,\quad k^l_{(ij)}=\delta^l_i x_c^j-\delta^l_j
x_c^i\,,\label{chichi}\\
&& k^0_{(04)}=-t_c\,,\quad k^i_{(04)}=-x_c^i\,.\nonumber
\end{eqnarray}
where we denote 
\begin{equation}
\chi=\frac{1}{2\omega}\left[1-\omega^2(t_c^2-{\vec{x}_c}^2)\right]\,.
\end{equation}
Taking into account that the particle of mass $m$ has the momentum $\vec{P}$  of components 
$P^i=-\omega m (k_{(0i)\,\mu} + k_{(i4)\,\mu})u_c^{\mu}$, 
we deduce the components of the four-velocity, 
\begin{eqnarray}
u_c^0&=&\frac{dt_c}{ds}=-\omega t_c \sqrt{1+\frac{\omega^2{P}^{2}}{m^2}t_c^2}\,,\nonumber\\
u_c^i&=&\frac{d{x_c^i}}{ds}=(\omega t_c)^2 \frac{P^i}{m}\,,\label{uu}
\end{eqnarray}
deriving the rectilinear timelike gedesic trajectory  \cite{CGRG},
\begin{equation}\label{geo}
\vec{x}_c(t)=\vec{x}_{c0}+\frac{\vec{P}}{\omega {P}^
2} \left(\sqrt{m^2+{P}^{2}\omega^2 t_{c0}^2}-\sqrt{ m^2+{P}^2
\omega^2 t_c^2}\, \right)\,,
\end{equation}
which is completely determined by the initial condition  $\vec{x}_c(t_{c0}) =\vec{x}_{c0}$  and the conserved momentum $\vec{P}$.  Consequently, the  conserved quantities  in an arbitrary point on geodesic, $(t_c,\vec{x}_c(t_c))$,  depend only on this point and the momentum $\vec{P}$   having the form  \cite{CGRG,CdS1},
\begin{eqnarray}
E&=&\omega\, \vec{x}_c(t_c)\cdot \vec{P}+\sqrt{ m^2+{P}^{2}\omega^2 t_c^2}\,,\label{Ene}\\
\vec{L}&=&\vec{x}_c(t_c)\land \vec{P}\,,\\
\vec{Q}&=&2\omega \vec{x}_c(t_c)E+\omega^2\vec{P}[t_c^2-\vec{x}_c(t_c)^2]\,.\label{Q}
\end{eqnarray}
These quantities are not independent since they satisfy,   
\begin{equation}
\vec{L}\cdot\vec{P}=\vec{L}\cdot\vec{Q}=0\,,\quad \vec{P}\land\vec{Q}=-2\omega E \vec{L} \,,
\end{equation}
which means that the vectors $\vec{x}_c(t_c)$, $\vec{P}$ and $\vec{Q}$ are in the same plane, orthogonal to $\vec{L}$. Moreover, one may verify the identity 
\begin{equation}\label{cas1}
E^2-\omega^2 {\vec{L}}^2-\vec{P}\cdot \vec{Q}=m^2\,,
\end{equation}
corresponding to the first Casimir operator of the $so(1,4)$ algebra \cite{CGRG}.    In the flat limit, $\omega\to 0$, when $-\omega t_c\to 1$ and $\vec{Q} \to \vec{P}$ this identity  becomes just  the usual mass-shell condition $p^2=m^2$ of special relativity. Thus we can conclude that there are only six independent conserved quantities, say the components of the vectors $(\vec{P},\vec{Q})$, that form a basis generating freely all the other conserved quantities.

On the other hand, we remark that the geodesic equation (\ref{geo}) can be split as $\vec{X}(t_c,\vec{x}_c)=\vec{X}(t_{c0},\vec{x}_{c0})$, for any arbitrary initial condition, if we introduce the new vector 
\begin{equation}\label{xi}
\vec{X}(t_c,\vec{x}_c)=\vec{x}_c(t_c) + \frac{\vec{P}}{\omega P^2}\sqrt{m^2+P^2\omega^2 t_c^2}\,,
\end{equation} 
which is an useful auxiliary conserved quantity that offers us the possibility of changing the basis $(\vec{P},\vec{Q})$ into the new one $(\vec{X},\vec{P})$ where we can write
\begin{eqnarray}
E&=&\omega \vec{X}\cdot \vec{P}\,,\\
\vec{L}&=&\vec{X}\land\vec{P}\,,\\
\vec{Q}&=&2\omega^2 (\vec{X}\cdot\vec{P}) \vec{X}-\left(\frac{m^2}{P^2}+\omega^2 {\vec{X}\,}^2\right)\vec{P}\,,
\end{eqnarray}
observing that the identity (\ref{cas1}) holds for any $\vec{X}$ and $\vec{P}$.

The conserved basis-vectors $(\vec{P},\vec{Q})$ or $(\vec{X},\vec{P})$ determine the plane of the geodesic trajectory where it is convenient to consider the Cartesian frame in $O$ whose orthonormal basis, $\{\vec{n}_{\perp},\vec{n}_P\}$, is formed by 
$\vec{n}_P=\frac{\vec{P}}{P}$ and its orthogonal complement, $\vec{n}_{\perp}$. In this frame we use the local Cartesian coordinates $(x_{\perp}, x_{\parallel})$ such that any position vector can be written as $\vec{x}=\vec{x}_{\perp}+\vec{x}_{\parallel}=x_{\perp} \vec{n}_{\perp}+x_{\parallel} \vec{n}_{P}$.   Notice that when $\vec{L}=0$ then  the geodesic is passing through the origin $O$ and, consequently, $\vec{n}_{\perp}$ remains undetermined.

\section{Geodesics in comoving charts}

Turning back to our problem of the relation among the initial conditions and the conserved quantities,  we study first first the comoving charts, i. e. the conformal, FLRW, and dSP ones. 

In the euclidean chart,  Eq.(\ref{Ene}) allows us to put the vector (\ref{xi})  in the form 
\begin{equation}\label{xi1}
\vec{X}(t_c,\vec{x}_c)=\vec{x}_{c\perp}+\frac{E}{\omega P}\,\vec{n}_P\,,
\end{equation} 
where $\vec{x}_{c\perp}=\vec{x}_c(t_c)-{\vec{n}_P}(\vec{x}_c(t_c)\cdot\vec{n}_P)$ is  the transverse part of $\vec{x}_c(t_c)$.  In other respects, from Eqs. (\ref{Q}) and (\ref{cas1}) we may write the useful representation,  
\begin{equation}\label{Qnp}
\vec{Q}=\vec{n}_{\perp}\frac{2\omega EL}{P}+\vec{n}_P\frac{E^2-m^2-\omega^2 L^2}{P}
\end{equation}
which helps us to find that the vector  
\begin{equation}\label{xperp}
\vec{x}_{c\perp}=\frac{1}{2\omega E}\,\vec{n}_{\perp}\cdot\vec{Q}=\frac{L}{P}\,\vec{n}_{\perp}=\vec{x}_{cA}\,,
\end{equation}
is conserved,  giving  the position of the particle at the time 
\begin{equation}
t_{c\perp}=- \frac{\sqrt{E^2-m^2}}{\omega P}\,,
\end{equation}
when this  is passing through the point $A$, of position vector $\vec{x}_c(t_{c\perp})=\vec{x}_{cA}$, where the energy (\ref{Ene}) takes the form
$E=\sqrt{m^2+\omega^2 P^2 t_{c\perp}^2}$.  

Now we can solve our problem by choosing the  initial condition $t_{c0}=t_{c\perp}$ and $\vec{x}_{c0}=\vec{x}_{cA}$ that allow us to write $\vec{X}(t_c,\vec{x}_c)=\vec{X}(t_{c\perp},\vec{x}_{cA}) $ deriving the geodesic trajectory  as
\begin{eqnarray}
\vec{x}_c(t_c)&=&\vec{x}_{cA}+\vec{n}_P\,{x}_{c\parallel}(t_c) \nonumber\\
& =&\vec{n}_{\perp}\frac{L}{P}+\vec{n}_P\frac{1}{\omega P}\left(E-\sqrt{m^2+P^2\omega^2 t_c^2}\right)\,.\label{geocon}
\end{eqnarray}
Thus we succeeded to express the geodesic equation in terms of conserved quantities without resorting to an explicit initial condition. The function ${x}_{c\parallel}(t_c)$ describes the motion along the direction $\vec{n}_P$ between the limits
\begin{equation}\label{dom1}
-\infty<{x}_{c\parallel}(t_c)\le \frac{E-m}{\omega P}\,,
\end{equation}
since $-\infty<t_c\le 0$.

{ \begin{figure}
  \centering
    \includegraphics[scale=0.53]{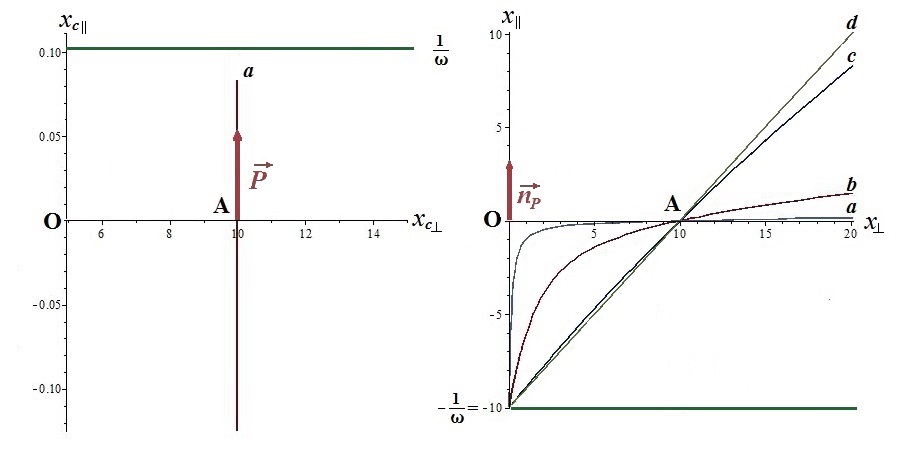}
    \caption{Geodesics in conformal coordinates (left) and dSP ones (right) for $\omega=0.1$, $t_{c\perp}=-\frac{1}{\omega}=-10$, $m=5$ and $P=0.1 (a)\,, 1(b),\, 10(c),\,100(d)$.   }
  \end{figure}}

The above results can be extended to any comoving chart with Cartesian space coordinates  the  time of which depends on $t_c$ but remaining independent on $\vec{x}_c$. The best example is the dSP chart $\{t,\vec{x}\}$ whose coordinates can be introduced  directly by substituting
\begin{equation}\label{EdS}
t_c=-\frac{1}{\omega}e^{-\omega t}\,, \quad \vec{x}_c=\vec{x}e^{-\omega t}\,,
\end{equation}
where $t\in(-\infty, \infty)$ is the {proper} time while $x^i$ are the 'physical' Cartesian space coordinates. This chart, having the line element 
\begin{equation}
ds^2=(1-\omega^2 {\vec{x}}^2)dt^2+2\omega \vec{x}\cdot d\vec{x}\,dt -d\vec{x}\cdot d\vec{x}\,, 
\end{equation}
is useful in applications since in the flat limit (when $\omega \to 0$) its coordinates become just the Cartesian ones of the Minkowski spacetime.  Performing then the substitution (\ref{EdS}) in Eq. (\ref{geocon}) we obtain the geodesic equation, 
\begin{eqnarray}
\vec{x}(t)&=&\vec{n}_{\perp}\,{x}_{\perp}(t)+\vec{n}_P\, x_{\parallel}(t) \nonumber\\
&=&\vec{n}_{\perp}\frac{L}{P}\,e^{\omega t}+\vec{n}_P \frac{1}{\omega P}\left(Ee^{\omega t}-\sqrt{m^2e^{2\omega t}+P^2}\right)\,,\label{xxpp}
\end{eqnarray}
whose transverse part is no longer conserved. Since in this chart $t\in {\mathbb R}$ the space domain of the geodesic is given by
\begin{equation}
0\le x_{\perp}(t)<\infty\,,\quad  -\frac{1}{\omega}\le x_{\parallel}(t)<\infty\,.
\end{equation}
The mobile reaches the point $A$, of coordinates $({x}_{A},0)$, at time $t_{\perp}$ which means that $x_A={x}_{\perp}(t_{\perp})$. Therefore, we find
\begin{eqnarray}
t_{\perp}&=&-\frac{1}{\omega}\ln (-\omega t_{c\perp})=-\frac{1}{\omega}\ln \frac{\sqrt{E^2-m^2}}{P}\,,\label{tperp}\\
x_A&=&-\frac{{x}_{cA}}{\omega t_{c\perp}}=\frac{L}{\sqrt{E^2-m^2}}\,.\label{xA2}
\end{eqnarray}   

Other chart frequently used in applications  is the FLRW one which combine the proper time with the conformal space coordinates, $\{t,\vec{x}_c\}$. In this chart the geodesic equation reads
\begin{equation}\label{FLRW}
\vec{x}_c(t)=\vec{n}_{\perp}\frac{L}{P}+\vec{n}_P\frac{1}{\omega P}\left(E-\sqrt{m^2+P^2e^{-2\omega t}}\right)\,,
\end{equation}
determining a rectilinear trajectory as in the conformal charts. The space domain remains the same as that of the conformal chart, given by  Eq. (\ref{dom1}), while point $A$ of coordinates $(t_{\perp},x_A)$ is reached now at proper time  (\ref{tperp}).  

The conclusion is that in comoving charts the geodesic trajectories are rectilinear only in the charts with conformal space coordinates, namely the conformal and FLRW ones. In the dSP charts these trajectories are, in general, curvilinear as in the right panel of Fig. 
1,  becoming rectilinear, along the momentum direction,  only  when $A=O$ since then ${L}=0$. Otherwise, the geodesic trajectory  is approaching to a rectilinear one only  in the ultra-relativistic regime, for $P\gg m$. An example is  the geodesic $d$ on the right panel of Fig. 1.

\section{Geodesics in static charts}

The above investigation cannot be extended to other types of charts since the geodesic equations $\vec{x}=\vec{x}(t)$ are not invariant under general diffeomorphisms involving simultaneously the time and space coordinates.  Therefore we must complete our study considering, in addition, parametric geodesic equations. 

Let us first turn back to the conformal chart where we introduce the new parameter
\begin{equation}
\lambda=\frac{1}{E}\sqrt{m^2+\omega^2P^2 t_c^2}\,>\frac{m}{E}\,,
\end{equation}
which increases monotonously  when $t_c$ decreases. With its help we derive the parametric geodesic equations,
\begin{eqnarray}
t_c(\lambda)&=&-\frac{1}{\omega P}\sqrt{E^2\lambda^2-m^2}\,,\\
\vec{x}_c(\lambda)&=&\vec{n}_{c\perp}\frac{L}{P}+\vec{n}_P\frac{E}{\omega P}(1-\lambda)\,,
\end{eqnarray} 
observing that the point $A$ correspond to the value $\lambda_A=1$ since $ t_c(\lambda)|_{\lambda=1}=t_{c\perp}$ and $\vec{x}_c(\lambda)|_{\lambda=1}=\vec{x}_{cA}$.

Similarly, we obtain the geodesic equations  in the dSP coordinates  
\begin{eqnarray}
t(\lambda)&=&-\frac{1}{\omega}\ln \frac{\sqrt{E^2\lambda^2-m^2}}{P}\,,\\
\vec{x}(\lambda)&=&\vec{n}_{c\perp}\frac{L}{\sqrt{E^2\lambda^2-m^2}}+\vec{n}_P\frac{E(1-\lambda)}{\omega \sqrt{E^2\lambda^2-m^2}}\,,\label{xdSP}
\end{eqnarray} 
as it results by inverting Eqs. (\ref{EdS}).

However, apart from the above simple examples our principal goal here is to obtain the geodesic equations in the static charts with the static time
\begin{equation}\label{ts}
t_s=-\frac{1}{2\omega}\ln\, \omega^2(t_c^2-\vec{x}_c^2)\,,
\end{equation}
and different types of Cartesian space coordinates. Using the same parameter $\lambda$ and  Eqs, (\ref{Ene}), (\ref{Q}) and (\ref{cas1}) we find the following parametric equation
\begin{equation}\label{tss}
t_s(\lambda)=-\frac{1}{2\omega}\ln \frac{2\lambda E^2-(E^2+m^2+\omega^2 L^2)}{P^2}\,,
\end{equation}
and the static time
\begin{equation}
t_{s\perp}=t_s(\lambda)|_{\lambda=1}=-\frac{1}{2\omega}\ln \frac{ E^2-m^2-\omega^2 L^2}{P^2}\,,
\end{equation}
when the particle reaches the point $A$.

The usual static chart $\{t_s, \vec{x}\}$ has the dSP Cartesian space coordinates and the line element, 
\begin{equation}
ds^2= (1-\omega^2 \vec{x}^2) dt_s^{2}-\left[\delta_{ij}+\omega^2 \frac{x^{i}x^{j}}{1-\omega^2 \vec{x}^2}\right]dx^{i}dx^{j}\,.\label{line3}
\end{equation} 
In this chart the parametric geodesic equations are (\ref{tss}) and (\ref{xdSP}) giving similar trajectories as in the right panel of Fig. 1.  Moreover, the parameter $\lambda$ can be eliminated but the closed form we obtain is complicated being useless in current applications.

For this reason, we look for other Cartesian coordinates in which the geodesic equations become simpler.  We observe that there is an useful identity
\begin{equation}\label{ident2}
1-\omega^2 \vec{x}^2=\frac{P^2}{E^2\lambda^2-m^2}e^{-2\omega t_s}\,,
\end{equation}  
which suggests us to use the {\em special} Cartesian coordinates of Refs. \cite{P1,P2} that read
\begin{equation}\label{xx3}
\vec{x}_s=\frac{\vec{x}}{\sqrt{1-\omega^2 \vec{x}^2}}\,,
\end{equation} 
giving the new line element
\begin{equation}
ds^2= \frac{1}{1+\omega^2 \vec{x}_s^2}\left[ dt_s^{2}-\left(\delta_{ij}-\omega^2 \frac{x_s^{i}x_s^{j}}{1+\omega^2 \vec{x}_s^2}\right)dx_s^{i}dx_s^{j}\right]\,.\label{line4}
\end{equation} 
In this chart the position of the point $A$ is given by the coordinates $(x_{sA},0)$ where
\begin{equation}
x_{sA}=\frac{L}{\sqrt{E^2-m^2-\omega^2L^2}}\,,
\end{equation}
as it results from Eqs. (\ref{xA2}) and (\ref{xx3}).

 { \begin{figure}
  \centering
    \includegraphics[scale=0.53]{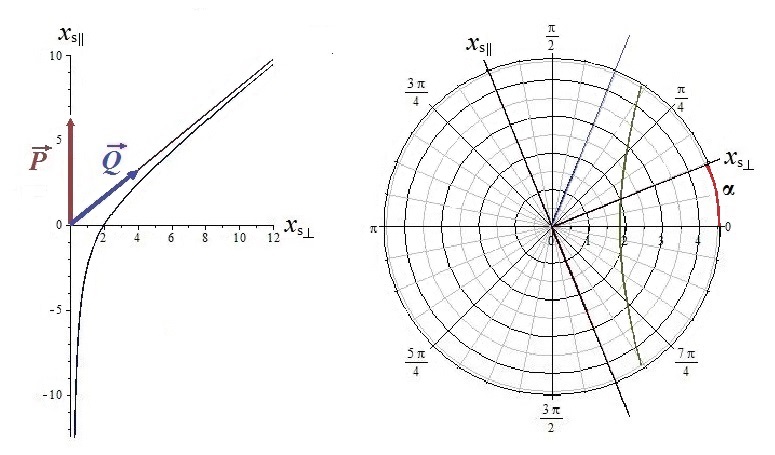}
    \caption{The geodesic in the special static chart for $\omega=0.1$ and the initial condition $t_{s0}=0$ is a hyperbola whose  principal axes are rotated with  the angle $\alpha=\frac{1}{2}{\rm angle}(\vec{P},\vec{Q})=-\frac{\pi}{8}$ with respect to the axes $(x_{s\perp},x_{s\parallel})$.} 
  \end{figure}}

The final task is to put the geodesic equation in a closed form eliminating the parameter $\lambda$ that can be deduced from Eq. (\ref{tss}) as 
\begin{equation}
\lambda=1+\frac{1}{2E^2}\left(P^2 e^{-2\omega t_s}-E^2+m^2+\omega^2 L^2\right)\,.
\end{equation}
Then, according to Eqs.  (\ref{xdSP}) and (\ref{ident2}),  we obtain the desired form of the geodesic equation in the chart $\{t_s,\vec{x}_s\}$ that reads
\begin{equation}\label{stsp}
\vec{x}_s(t_s)=\vec{n}_{\perp}\frac{L}{P}\,e^{\omega t_s}+\vec{n}_P \frac{(E^2-m^2-\omega^2L^2)e^{\omega t_s}-P^2e^{-\omega t_s}}{2\omega E P}\,.
\end{equation}   
This result is remarkable since  Eq. (\ref{Qnp}) allows us to rewrite it simply as 
\begin{equation}\label{geoPQ}
\vec{x}_s(t_s)=\frac{1}{2\omega E}\left(\vec{Q}e^{\omega t_s}-\vec{P}e^{-\omega t_s}\right)\,,
\end{equation}
or in the equivalent form, 
\begin{equation}
\vec{x}_s(t_s)=\frac{1}{E}\left(\vec{K}\cosh \omega t_s-\vec{R}\sinh\omega t_s\right)\,,
\end{equation}
resulted from Eqs. (\ref{PQKR}). We obtain thus the Cartesian version of our previous result obtained recently in spherical coordinates \cite{CdS2} according to which the geodesics in the chart $\{t_s,\vec{x}_s\}$ are hyperbolas whose asymptotes are in the directions of $-\vec{P}$ and $\vec{Q}$ as in Fig. 2.

\section{Null cones and horizons}

On dS spacetime the null (or light) cones are important since the Killing vector $k_{(04)}$ giving the conserved energy is time-like only inside the null cone. Outside the light cone this can be space-like which means that energy cannot be correctly defined in this domain \cite{CGRG}. However this is not an impediment since the physical observation can be done only inside the null cones where we meet the timelike worldines of massive particles  or on the null cone which is defined by the wordlines of massless particles. In Fig. 3 we give the  example of  wordlines of a massive and a massless particle in the static and special static charts.

 { \begin{figure}
  \centering
    \includegraphics[scale=0.53]{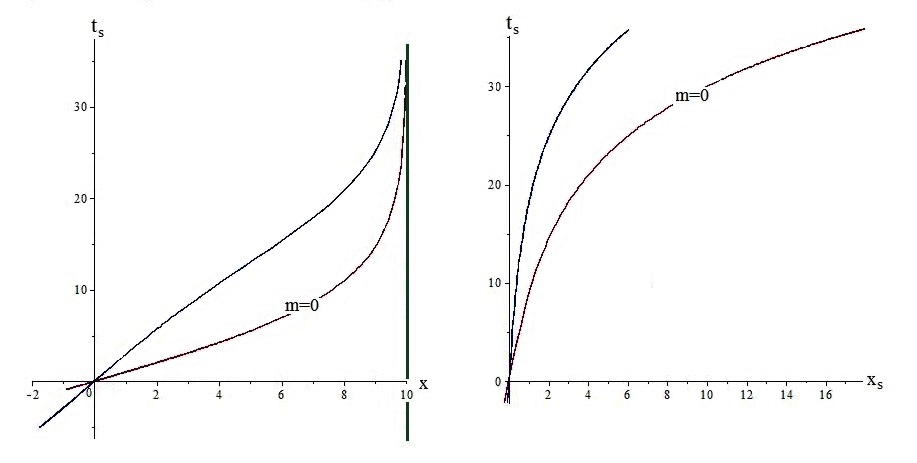}
    \caption{Worldilines of massive and massless particles in the static chart (left) and  special static one  (right) for $\omega=0.1$ and the initial condition $t_{s0}=0$. In the static chart there is an event horizon at $x=-\frac{1}{\omega}$.}
  \end{figure}} 

For analysing the form of the null cones, we focus first on the null geodesics setting $t_{c\perp}=-\frac{1}{\omega}$ since then the energy takes the familiar form of special relativity $E=\sqrt{m^2+P^2}$. In this case the initial conditions in the charts we considered here become\begin{eqnarray}
\begin{array}{l}
t_{c0}=t_{c\perp}=\textstyle{-\frac{\textstyle 1}{\textstyle \omega}}\\
~~\\
\vec{x}_{c0}=\vec{x}_{c\perp}=\frac{\textstyle L}{\textstyle P}\,\vec{n}_{\perp}
\end{array} &\to&
\begin{array}{l}
t_{0}=t_{\perp}=0\\
~~\\
\vec{x}_{0}=\vec{x}_{A}=\frac{\textstyle L}{\textstyle P}\,\vec{n}_{\perp}
\end{array}\nonumber\\
& \to&
\begin{array}{l}
t_{s0}=t_{s\perp}=-\frac{\textstyle \textstyle 1}{\textstyle  2\omega}\ln\left(1-\frac{\textstyle \omega^2 L^2}{\textstyle P^2}\right)\\
~~\\
\vec{x}_{s0}=\vec{x}_{sA}=\frac{\textstyle L}{\textstyle \sqrt{E^2-m^2-\omega^2L^2}}\,\vec{n}_{\perp}
\end{array} 
\end{eqnarray}
with $\vec{x}_{c0}=\vec{x}_{0}$ (as in Fig. 1).

Under such circumstances,  the null geodesics with $m=0$ but arbitrary $L\not= 0$ can be written as
\begin{eqnarray}
{\rm Chart~~~~}&&\rm{Null~ geosesic}\\
{\rm conform.} &&\vec{x}_c(t_c)=\vec{n}_{\perp}\frac{L}{P}+\vec{n}_P \frac{1+\omega t_c}{\omega}\,,  \\
{\rm FLRW~~} &&\vec{x}_c(t)=\vec{n}_{\perp}\frac{L}{P}+\vec{n}_P \frac{1-e^{-\omega t}}{\omega}\,,  \\
{\rm dSP~~~~~}&&\vec{x}(t)=\vec{n}_{\perp}\frac{L}{P}\,e^{\omega t}+\vec{n}_P \frac{e^{\omega t}-1}{\omega}\,,\\
{\rm stat.~ sp.}&& \vec{x}_s(t_s)=\vec{n}_{\perp}\frac{L}{P}\,e^{\omega t_s}+\vec{n}_P\frac{1}{2\omega}\left[\left(1-\frac{\omega^2 L^2}{P^2}\right)e^{\omega t_s}-e^{-\omega t_s}\right]\,.
\end{eqnarray}
Notice that these equations are obtained directly from the geodesc equations in closed form (\ref{geocon}), (\ref{xxpp}), (\ref{FLRW}) and (\ref{stsp})  apart from the static chart where we have to use the parametric equations Eqs. (\ref{tss}) and (\ref{xdSP}) that for $m=0$ yield
\begin{eqnarray}
t_s(\lambda)&=&-\frac{1}{2\omega}\ln \left(2\lambda-1-\frac{\omega^2 L^2}{P^2}\right)\,,\\
\vec{x}(\lambda)&=&\vec{n}_{\perp} \frac{L}{\lambda P}+\vec{n}_P\frac{1-\lambda}{\omega\lambda}\,.
\end{eqnarray}

Hereby we may deduce the form of the null cones in the charts under consideration focusing only on the rectilinear geodesics that are passing through origin, having $\vec{L}=0$, $\vec{Q}=\vec{P}$  and  standard initial conditions
\begin{eqnarray}
t_{c0}=-\frac{1}{\omega}& \to& t_0=t_{s0}=0\,,\\
&&\vec{x}_{c0}=\vec{x}_{0}=\vec{x}_{s0}=0\,.
\end{eqnarray}
Then we find that the corresponding worldlines have the following equations   
\begin{center}
\begin{tabular}{llll}
Chart&Worldline (right)&Domain (x)&Domain (t)\\
&&&\\
conform.&$t_c= x_c-\frac{1}{\omega}$&$x_c\in (-\infty, \frac{1}{\omega}]$&$t_c\in(-\infty, 0]$\\
&&&\\
FLRW&$t=-\frac{1}{\omega}\ln (1-\omega x_c)$&$x_c\in (-\infty, \frac{1}{\omega}]$&$t\in{\mathbb R}$\\
&&&\\
dSP&$t=\frac{1}{\omega}\ln (1+\omega x)$&$x\in [-\frac{1}{\omega},\infty)$&$t\in{\mathbb R}$ \\
&&&\\
static&$ t_s=\frac{1}{\omega}\,{\rm arctanh}\,\omega x$&$x\in [-\frac{1}{\omega}, \frac{1}{\omega}]$&$t_s\in{\mathbb R}$ \\
&&&\\
static&$ t_s=\frac{1}{\omega}\,{\rm arcsinh}\,\omega x_s$&$x_s\in{\mathbb R}$&$t_s\in{\mathbb R}$\\
special&&&\\
\end{tabular}
\end{center}
where $x_c$, $x$ and $x_s$ are space coordinates along the geodesic direction $\vec{x}_P$. These equations give only one of the intersections of the null cone with the plane $(t,x)$ while the second one can be obtained by applying the parity transformation along the direction $\vec{n}_P$.

{ \begin{figure}
  \centering
    \includegraphics[scale=0.53]{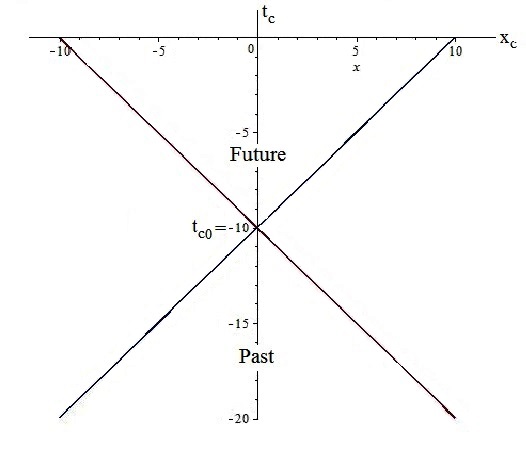}
    \caption{Null cone of the conformal chart for the initial condition $t_{c0}=t_{c\perp}=-\frac{1}{\omega}=-10$.}
  \end{figure}} 

{ \begin{figure}
  \centering
    \includegraphics[scale=0.50]{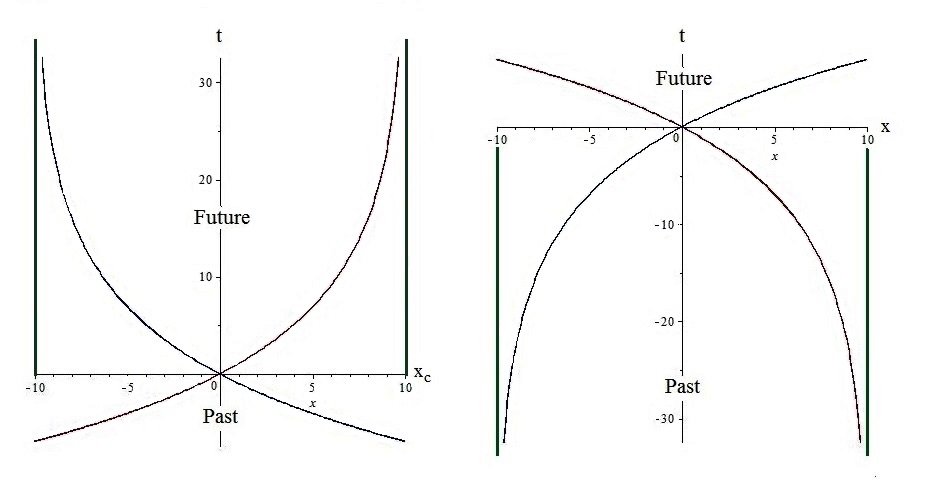}
    \caption{Null cones of the FLRW chart (left) and dSP one (right) for $\omega=0.1$ and the initial condition $t_0=0$.}
  \end{figure}} 

{ \begin{figure}
  \centering
    \includegraphics[scale=0.50]{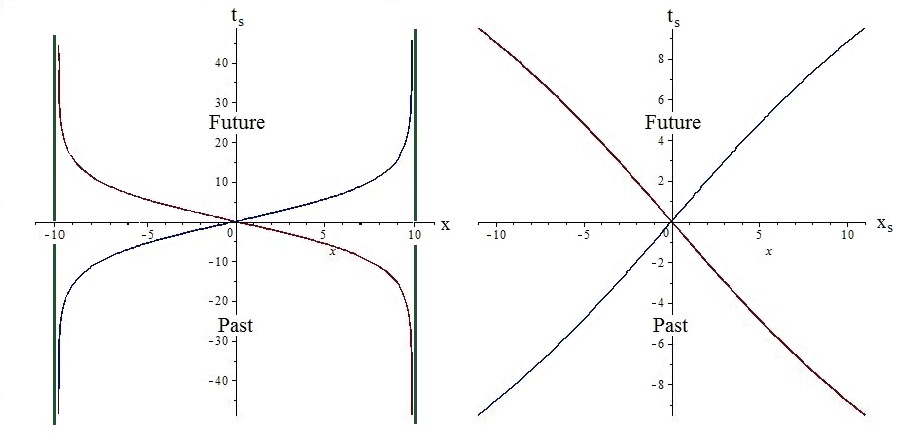}
    \caption{Null cones of the static chart (left) and special static one (right) for $\omega=0.1$ and the initial condition $t_{s0}=0$.}
  \end{figure}} 
  
The null cones in different charts have specific forms depending on the coordinates we use as in Figs. 4-6. As observed before, the dSP chart is the 'physical' one whose space coordinates measure the physical distances. In this chart we observe the event horizon of radius $\frac{1}{\omega}$ which is the limit of the space domain from which the past events can be observed \cite{BD}. On the static chart, the condition $|\vec{x}|\le\frac{1}{\omega}$ gives an event horizon  and, in addition,  restricts the future motions up to a border which plays the role of an 'expectation' horizon (Fig. 6 left), indicating the limit that can be reached by a light beam emitted in origin when $t\to \infty$.  Another horizon of this type we meet in the case of the FLRW charts (Fig. 5 left). 

\section{Concluding remarks}

We studied the geodesic equations in terms of conserved quantities using exclusively Cartesian space coordinates $\vec{x}\propto \vec{z}$ since in this case the $SO(3)$ symmetry becomes global such that  the Cartesian coordinates and all the conserved vectors we met here transform alike under rotations. 

The principal advantage of this choice is that we can write the geodesic equations  in  intuitive simple forms as, for example, Eq. (\ref{geoPQ}) which lays out explicitly the positions of the symptotes of the geodesic trajectory in the special static charts. 

Technically speaking, the global $SO(3)$ symmetry helps us to introduce associated spherical coordinates, $\vec{x}\to (r, \theta, \phi)$,  with $r=|\vec{x}|$, whose angular variables are the same in all the charts we met here.  This means that the changes of variables will involve only the pair $(t,r)$ which, according to Eq. (\ref{trcor}), transform as  
 \begin{eqnarray}\label{trcor1}
t_c&=&-\frac{1}{\omega}\,e^{-\omega t}=-\frac{1}{\omega}\sqrt{1+\omega^2{r}_s^2}\,e^{-\omega t_s}\,,\\
{r}_c&=&{r}e^{-\omega t}={r}_s e^{-\omega t_s}\,.
\end{eqnarray} 

Another advantage is that in all these charts we may consider the same 3-dimensional othonormal Cartesian basis as in  $M^5$, which will help us to control the position of the plane $(\vec{P},\vec{Q})$ of the geodesic trajectory.  We specify that this basis must not be confused with that of the tetrad vector fields whose orthogonality is defined with respect the dS metric.  

Finally, we note that all the new results presented here were obtained without using the covariant formalism of general relativity. In other words, when we have conserved quantities we can proceed as in classical mechanics, exploiting prime integrals instead of integrating geodesic equations in covariant form. 

\subsection*{Acknowledgements}

This work was supported by a grant of the Ministry of National Education and Scientific Research, RDI Programme for Space Technology and Advanced Research - STAR, project number 181/20.07.2017.

\end{document}